\documentclass[12pt,tightenlines,showpacs,preprintnumbers,%
superscriptaddress,nofootinbib]{revtex4}
\newcommand{\PRE}[1]{{#1}} 
\usepackage{bm}
\usepackage{subfigure}
\usepackage{slashed}
\usepackage{epsfig}
\usepackage{hyperref}

\def\beq{\begin{eqnarray}}
\def\eeq{\end{eqnarray}}
\def\bea{\begin{eqnarray}}
\def\eea{\end{eqnarray}}

\newcommand{\mev}{\text{MeV}}
\newcommand{\gev}{\text{GeV}}

\newcommand{\eqref}[1]{Eq.~(\ref{#1})}

\newcommand{\secref}[1]{Sec.~\ref{sec:#1}}

\newcommand{\figref}[1]{Fig.~\ref{fig:#1}}

\newcommand{\gsim}{\lower.7ex\hbox{$\;\stackrel{\textstyle>}{\sim}\;$}}
\newcommand{\lsim}{\lower.7ex\hbox{$\;\stackrel{\textstyle<}{\sim}\;$}}

\begin{document}

\preprint{UCI-TR-2013-11, UH511-1211-2013, CALT 68-2938}

\title{ \PRE{\vspace*{1.5in}}
Xenophobic Dark Matter
\PRE{\vspace*{0.3in}} }

\author{Jonathan L.~Feng}
\email{jlf@uci.edu}
\affiliation{Department of Physics and Astronomy, University of
California, Irvine, CA 92697, USA
\PRE{\vspace*{.2in}}
}

\author{Jason Kumar}
\email{jkumar@hawaii.edu}
\affiliation{Department of Physics and Astronomy, University of
Hawaii, Honolulu, HI 96822, USA
\PRE{\vspace*{.2in}}
}

\author{David Sanford \PRE{\vspace*{.3in}}}
\email{dsanford@caltech.edu}
\affiliation{California Institute of Technology, Pasadena, CA 91125,
  USA
\PRE{\vspace*{.5in}}
}

\begin{abstract}
\PRE{\vspace*{.3in}} We consider models of xenophobic dark matter, in
which isospin-violating dark matter-nucleon interactions significantly
degrade the response of xenon direct detection experiments. For models
of near-maximal xenophobia, with neutron-to-proton coupling ratio $f_n
/ f_p \approx -0.64$, and dark matter mass near 8 GeV, the regions of
interest for CoGeNT and CDMS-Si and the region of interest identified
by Collar and Fields in CDMS-Ge data can be brought into agreement.
This model may be tested in future direct, indirect, and collider
searches.  Interestingly, because the natural isotope abundance of
xenon implies that xenophobia has its limits, we find that this
xenophobic model may be probed in the near future by xenon
experiments.  Near-future data from the LHC and Fermi-LAT may also
provide interesting alternative probes of xenophobic dark matter.
\end{abstract}


\maketitle

\newpage

\section{Introduction}

\begin{figure}[tb]
\subfigure[]{
\includegraphics*[width=0.31\textwidth]{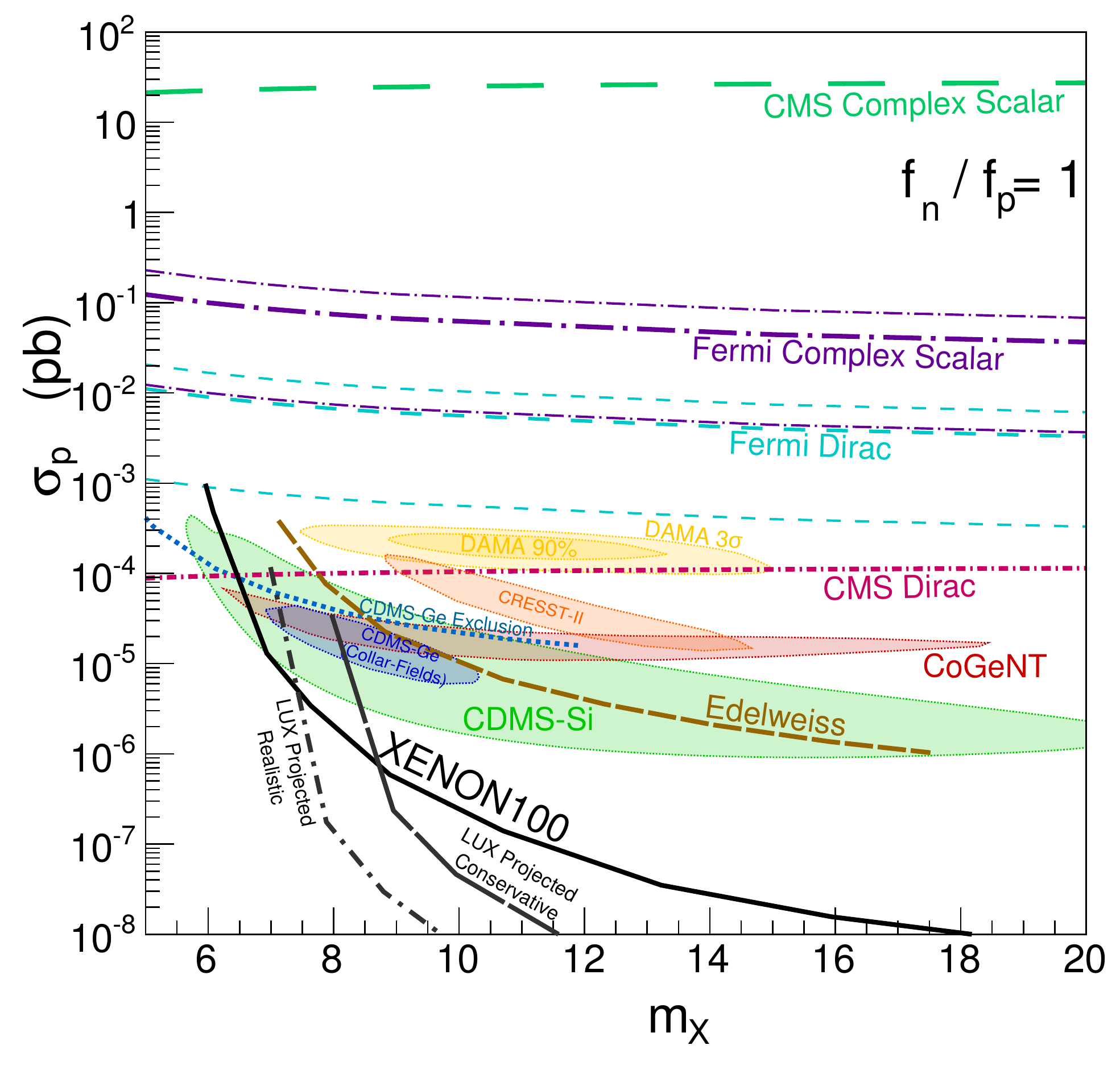}
\label{fig:sigmap_1}}
\subfigure[]{
\includegraphics*[width=0.31\textwidth]{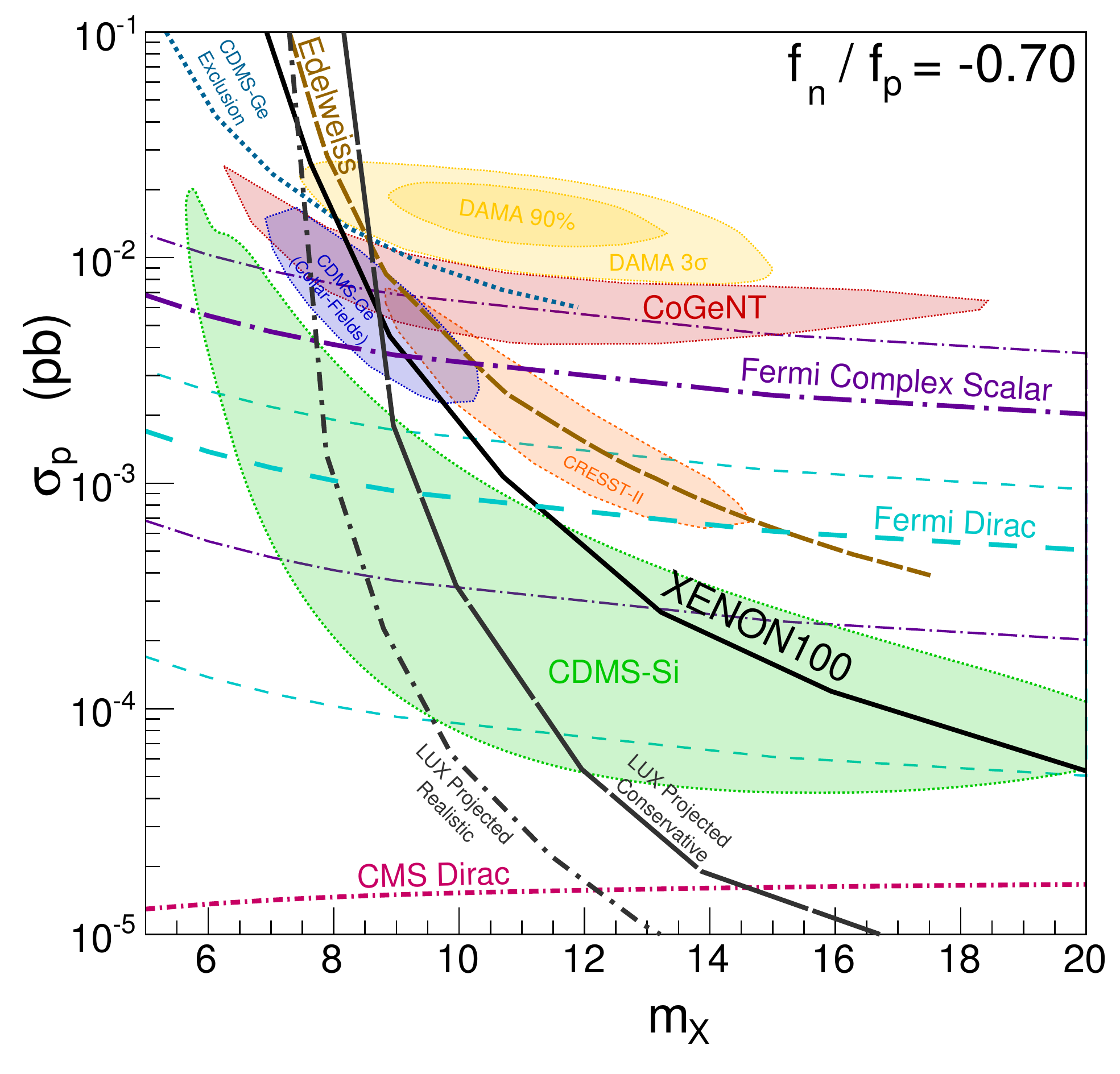}
\label{fig:sigmap_-7}}
\subfigure[]{
  \includegraphics*[width=0.31\textwidth]{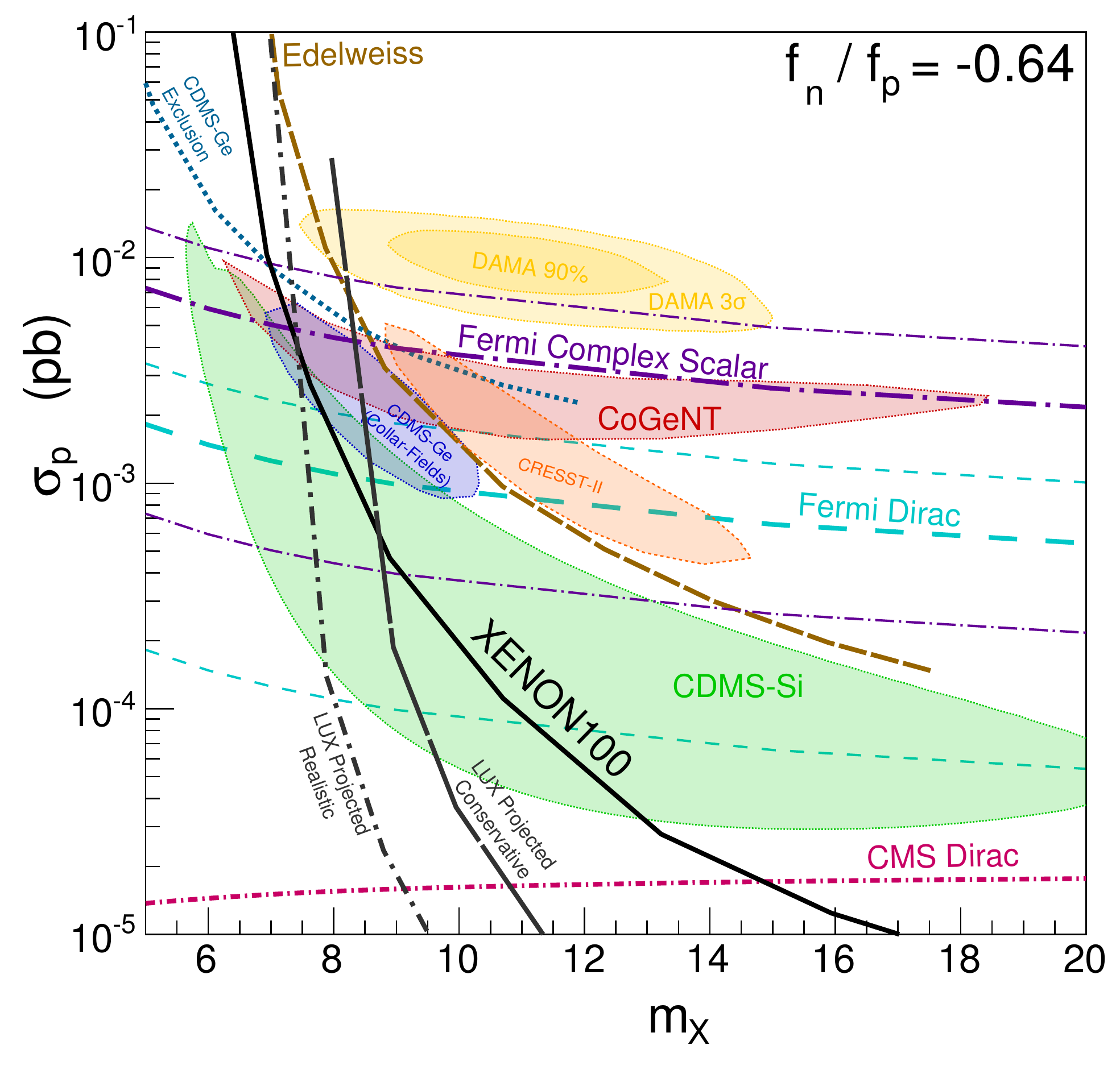}
\label{fig:sigmap_-64}}
\vspace*{-.1in}
\caption{\label{fig:sigmap_onepoint} \textit{Regions of interest and
    exclusion contours in the $(m_X, \sigma_p)$ for neutron-to-proton
    coupling ratios $f_n / f_p = 1$ (left), $f_n / f_p = -0.70$
    (center) and $f_n / f_p = -0.64$ (right).}  Plotted are the 90\%
  CL ROIs for CDMS-Si~\cite{Agnese:2013rvf},
  CoGeNT~\cite{Aalseth:2010vx}, and CDMS-Ge
  (Collar/Fields)~\cite{Collar:2012ed}, the 90\% and $3\sigma$ ROIs
  for DAMA~\cite{Bernabei:2010mq} as determined in
  Refs.~\cite{Savage:2008er, Savage:2010tg}, and exclusion contours
  from XENON100~\cite{Aprile:2011hi,Aprile:2012nq},
  Edelweiss~\cite{Armengaud:2012pfa}, and CDMS~\cite{Ahmed:2010wy},
  and projected exclusion bounds from LUX~\cite{Akerib:2012ak}.  Also
  plotted are 90\% CL exclusion contours from CMS and from the
  Fermi-LAT, assuming dark matter is either a complex scalar or Dirac
  fermion coupling only to first generation quarks through an
  effective contact interaction permitting unsuppressed
  spin-independent scattering and $S$-wave annihilation.  The thin
  dot-dashed violet and dashed teal lines correspond to the systematic
  uncertainty in the Fermi-LAT bounds from astrophysical uncertainties
  for complex scalar and Dirac fermion candidates, respectively.  In
  the center and right panels the CMS Complex Scalar exclusion bounds
  exceed the plotted range by between one and two orders of magnitude,
  and thus place no constraints on the disputed region.}
\end{figure}

In recent years experimental progress in direct detection of dark
matter has been extraordinary, especially for dark matter that
exhibits spin-independent (SI) nuclear scattering.  Candidates with
relatively low masses of $\sim 10$ GeV have been particularly
exciting, with potential signals at DAMA~\cite{Bernabei:2010mq},
CoGeNT~\cite{Aalseth:2010vx}, CRESST~\cite{Angloher:2011uu}, and
CDMS~\cite{Agnese:2013rvf}.  Although the tentative signals are
generally within the same region of parameter space, they do not
produce consistent determinations of either mass or interaction cross
section given conventional assumptions.  Moreover, several direct
detection experiments have reported the absence of an excess of
events, with XENON100~\cite{Aprile:2011hi,Aprile:2012nq} placing
particularly strong constraints on these results (see
\figref{sigmap_1}). This has led to recent attempts to reconcile the
results of these experiments by considering theories that deviate from
standard assumptions about dark matter interactions or its
astrophysical
distributions~\cite{Frandsen:2013cna,Mao:2013nda,Cotta:2013jna}.

In this work, we focus on the possibility of isospin-violating dark
matter (IVDM), in which dark matter interacts with protons and
neutrons with different couplings~\cite{Kurylov:2003ra,%
  Giuliani:2005my,Chang:2010yk,Kang:2010mh,Feng:2011vu}.  IVDM is a
highly motivated generalization of the conventional isospin-invariant
case: weakly-interacting massive particles (WIMPs) do not resolve the
internal structure of nucleons, but they do resolve the nucleon
structure within nuclei.  Irrespective of attempts to explain or
reconcile data, IVDM parametrizes the scattering off of matter in
terms of the smallest structure WIMP scattering resolves.  Indeed, in
the spin-dependent direct detection literature, isospin-violating
effects are generally considered.  Although some well-known dark
matter candidates, such as the neutralino in simple supersymmetric
models, have effectively isospin-invariant interactions, this is not
generically the case, as we detail below.  In this data-rich era, it
is appropriate to shed theoretical prejudices to the extent possible,
and IVDM provides an extremely natural framework to analyze direct
detection data.

A re-analysis of IVDM is motivated by several recent developments in
direct detection experiments, including limits from the
XENON10~\cite{Angle:2011th} and
XENON100~\cite{Aprile:2011hi,Aprile:2012nq} experiments, new exclusion
contours from a low-mass analysis of CDMS-Ge
detectors~\cite{Akerib:2010pv,Ahmed:2010wy}, modifications to the
CoGeNT region of interest (ROI)~\cite{Aalseth:2012if,Kelso:2011gd} due
to greater exposure and a better understanding of surface event
contamination, a new ROI arising from an excess of events seen by
CRESST~\cite{Angloher:2011uu}, and most recently a new ROI arising
from an excess of events seen by the CDMS-Si
detectors~\cite{Agnese:2013rvf}.

In light of these developments, this paper will revisit IVDM in the
context of low-mass dark matter.  The tightest constraints on low-mass
dark matter arise from XENON100, so any attempt to reconcile the
positive signals of some detectors with the negative signals from
others must focus on {\em xenophobic dark matter}, in which the
sensitivity of xenon-based detectors is highly suppressed by
destructive interference between proton and neutron interactions.  It
is important to note~\cite{Feng:2011vu} that the sensitivity of
xenon-based detectors cannot be suppressed arbitrarily, given the
significant abundances of multiple isotopes of xenon; no choice of
relative couplings can completely cancel the response of all of
xenon's isotopes simultaneously.  For example, it does not appear
possible to obtain consistency between XENON100 exclusion contours and
either the DAMA or CRESST ROIs, even with maximally xenophobic dark
matter.  As a result, we will not focus on those experiments.

We will take as our guideposts the CoGeNT ROI found in
Refs.~\cite{Aalseth:2012if,Kelso:2011gd} and the ROI found by Collar
and Fields in an analysis of the recoil spectrum of all CDMS-Ge
detectors~\cite{Collar:2012ed}.  Although there have been several
questions regarding the status of signals in germanium-based
detectors, we will find it useful to treat these ROIs as benchmarks,
because they identify a relatively small region of parameter space
near $m_X \sim 8~\gev$ in which the potential signals and exclusion
contours of current germanium-based detectors can all be satisfied.
Our focus will be on obtaining consistency of these regions with
exclusion contours from xenon-based detectors, consistency with the
CDMS-Si ROI, and ways of testing these models with near-future direct
detection, indirect detection, and collider searches.

We will find that although maximally xenophobic dark matter with $f_n
/ f_p \simeq -0.70$ (\figref{sigmap_-7}) reduces the tension between
the germanium-based ROIs and the xenon-based exclusion contours, the
germanium-based and silicon-based ROIs do not overlap.  On the other
hand, for near-maximally xenophobic dark matter with $f_n / f_p
=-0.64$ (\figref{sigmap_-64}), there is a region of parameter space
where the germanium-based and silicon-based ROIs overlap which is
consistent with xenon-based 90\% CL exclusion contours.  This model
may be decisively probed by the LUX experiment~\cite{Akerib:2012ak}.
Moreover, we will see that xenophobic dark matter is much more
amenable to indirect and collider detection strategies; some
xenophobic models for the low-mass data can be excluded by current CMS
monojet analyses, while other models will be tested soon with new
Fermi-LAT data on gamma rays from dwarf spheroidal galaxies.  More
generally we will see that, given any signals of dark matter at a
direct detection experiment, it is necessary to compare the results of
multiple experiments, including collider experiments and experiments
using indirect detection strategies, to determine the dark
matter-nucleon couplings.

In \secref{general} we review the general nature of isospin-violating
couplings and discuss the relationship between the
normalized-to-nucleon cross section usually reported by experiments
and the actual dark matter-proton scattering cross section.  In
\secref{xenophobic} we focus on xenophobic dark matter.  We conclude
with a discussion of our results in \secref{conclusions}.

\section{General Isospin-Violating Couplings}
\label{sec:general}

\subsection{The Case for Isospin-Violating Interactions}

Although isospin-invariant couplings are generally assumed when
reporting direct detection results, isospin-violating couplings are in
fact generic for theories with WIMPs.  This results from the fact that
interactions of WIMPs are typically related to electroweak symmetry,
and, in particular, to hypercharge.  Since right-handed up and down
quarks have different hypercharge, it would be natural to expect these
interactions to depend on isospin.  The fact that the spin-independent
scattering matrix element is largely isospin-invariant for WIMPs in
some scenarios, such as the constrained minimal supersymmetric
standard model (CMSSM), is actually the result of several non-trivial
coincidences.  For example, in the CMSSM, the Bino component of the
lightest supersymmetry particle (LSP) can scatter off nuclei through
squark exchange, and this matrix element is generally
isospin-violating, but the spin-independent piece of the matrix
element is proportional to the left-right squark-mixing angle.  Under
the assumption of minimal flavor violation (as in the CMSSM) this
angle is small for first generation squarks.  The Higgsino component
can also scatter through $Z$-exchange, which again produces an
isospin-violating contribution to the matrix element.  But since the
LSP is a Majorana fermion, the leading term is again spin-dependent.
The Wino/Higgsino component of LSP can scatter off nuclei through
Higgs exchange, and this contribution to the scattering matrix element
is spin-independent, but it is also largely isospin-invariant because
the coupling is proportional to the quark mass.  The assumption of
isospin-invariant interactions is really only justified within this
narrow framework and others like it, and these frameworks can realize
a low-mass dark matter candidate only with great difficulty.

Indeed, for many models of dark matter, couplings to nucleons are
indeed isospin-violating.  Dark matter in the form of a Dirac fermion
or complex scalar that is part of a weak doublet will naturally couple
in an isospin-violating manner because of the difference in
hypercharge of the up and down quarks.  This is also the case for
dark matter charged under a hidden U(1) gauge group with small kinetic
mixing with hypercharge.  Likewise, new scalar or fermionic mediators
generically couple in a flavor non-universal manner, which can produce
isospin-violating couplings to nucleons.

In considering a generic model of dark matter, and in particular a
model that could explain the low-mass data, one should really treat
the relative coupling to protons and neutrons as a free parameter that
can only be determined with guidance from the data.  This assumption
is sufficient for comparing direct detection experiments, as the
relative coupling to protons and neutrons completely define the
parameter space.  Further assumptions are required when comparing to
indirect detection and collider results, and, in particular,
assumptions about the flavor structure of the interaction are
required, since fixing the ratio of proton and neutron cross sections
does not uniquely specify the theory.  The type of candidate and
mediation mechanism structure also have a significant impact on the
comparison between direct detection results and indirect and collider
searches.

\subsection{Direct Detection}

If dark matter interacts with standard model matter through an elastic
contact interaction, then the spin-independent differential scattering
cross section can be written as
\begin{eqnarray}
{d \sigma \over dE_R} &=& {\mu_A^2 \over M_*^4} [f_p Z + f_n (A-Z)]^2
\left[{m_A \over 2\mu_A^2 v^2} F^2 (E_R) \right] ,
\end{eqnarray}
where $E_R$ is the recoil energy, $m_A$ is the mass of the target
nucleus, $\mu_A = m_X m_A / (m_X + m_A)$ is the reduced mass, and
$F(E_R)$ is a nuclear form factor (assumed to be the same for protons
and neutrons).  The couplings $f_p$ and $f_n$ parametrize the
strengths of dark matter coupling to protons and neutrons,
respectively; the interactions are isospin-invariant if $f_n = f_p$.
The rate of events at a direct detection experiment is thus
proportional to the zero-momentum transfer scattering cross section
\begin{eqnarray}
\hat \sigma_A & = & {\mu_A^2 \over M_*^4} [f_p Z + f_n (A-Z)]^2 \ ,
\end{eqnarray}
where the proportionality constant is independent of the particle
physics model, and is determined by the nuclear form factor, the
velocity distribution, the target size, and the energy threshold of
the experiment.

Direct detection experiments typically report results in terms of
$\sigma_N^Z$, a ``normalized-to-nucleon cross section.''  This is the
nucleon-dark matter scattering cross section that would be inferred,
assuming $f_n / f_p =1$, from the data of a detector using a target
with $Z$ protons.  For a given isotope with $Z$ protons and $A$
nucleons, the normalized-to-nucleon cross section is related to the
dark matter-nucleus zero-momentum transfer cross section by
$\sigma_N^{Z} = (\hat \sigma_A / A^2) \times (\mu_p^2 / \mu_A^2)$,
where $\mu_p$ is the dark matter-proton reduced mass.

If dark matter interactions are actually isospin-invariant, then
$\sigma_N^{Z}$ is equal to the proton-dark matter and neutron-dark
matter scattering cross sections $\sigma_p$ and $\sigma_n$.  For a
general ratio of couplings $f_n / f_p$, $\sigma_N^Z$ is related to
$\sigma_p$ and $\sigma_n$ by the ``degradation factors'' $D^Z_{p, n}$,
defined as\footnote{Note, $D_p^Z \equiv 1/F_Z$, where $F_Z$ is defined
  in Ref.~\cite{Feng:2011vu}.}
\begin{eqnarray}
D^Z_p \equiv \frac{\sigma_N^Z}{\sigma_p} &=& \frac{\sum_i \eta_i
  \mu_{A_i}^2 [Z + (f_n / f_p) (A_i -Z)]^2 }{\sum_i \eta_i \mu_{A_i}^2
  A_i^2} \\
D^Z_n \equiv \frac{\sigma_N^Z}{\sigma_n} &=& D^Z_p \left(
\frac{f_p}{f_n} \right)^2
\label{eq:FZ}
\end{eqnarray}
where the sum is over isotopes $i$, and $\eta_i$ is the natural
abundance of the $i$th isotope.  If a direct detection experiment uses
a target with $Z$ protons, then $D^Z_p$ is the reduction in
sensitivity to $\sigma_p$ relative to the isospin-invariant case, and
rescales the event rate expected for a given value of $\sigma_p$.  For
elements with only one naturally-abundant isotope, there exists a
choice of $f_n / f_p$ such that $D^Z_{p, n} \to 0$, resulting in zero
sensitivity for scattering off those elements.  In contrast, if an
element has multiple isotopes, there is a lower bound on $D^Z_{p, n}$,
since completely destructive interference cannot be simultaneously
achieved for all isotopes at once, and there is a reduced but non-zero
sensitivity as a worst-case scenario in such
elements~\cite{Feng:2011vu}.  An important caveat to these statements
is that NLO corrections, including loop corrections and multiparticle
exchange, can have a significant effect when the leading order
scattering cross section is suppressed~\cite{Cirigliano:2012pq}.  But
as the analysis of this effect is model-dependent, we will not
consider it further. We do note that Ref.~\cite{Cirigliano:2012pq}
found such effects could either reduce or increase the maximal value
of $D^Z_p$ in elements with multiple naturally-occurring isotopes.

\begin{figure}[tb]
\subfigure[\ $\sigma^N_Z / \sigma_p$, entire $f_n/f_p$ range]{
\includegraphics*[width=0.48\textwidth]{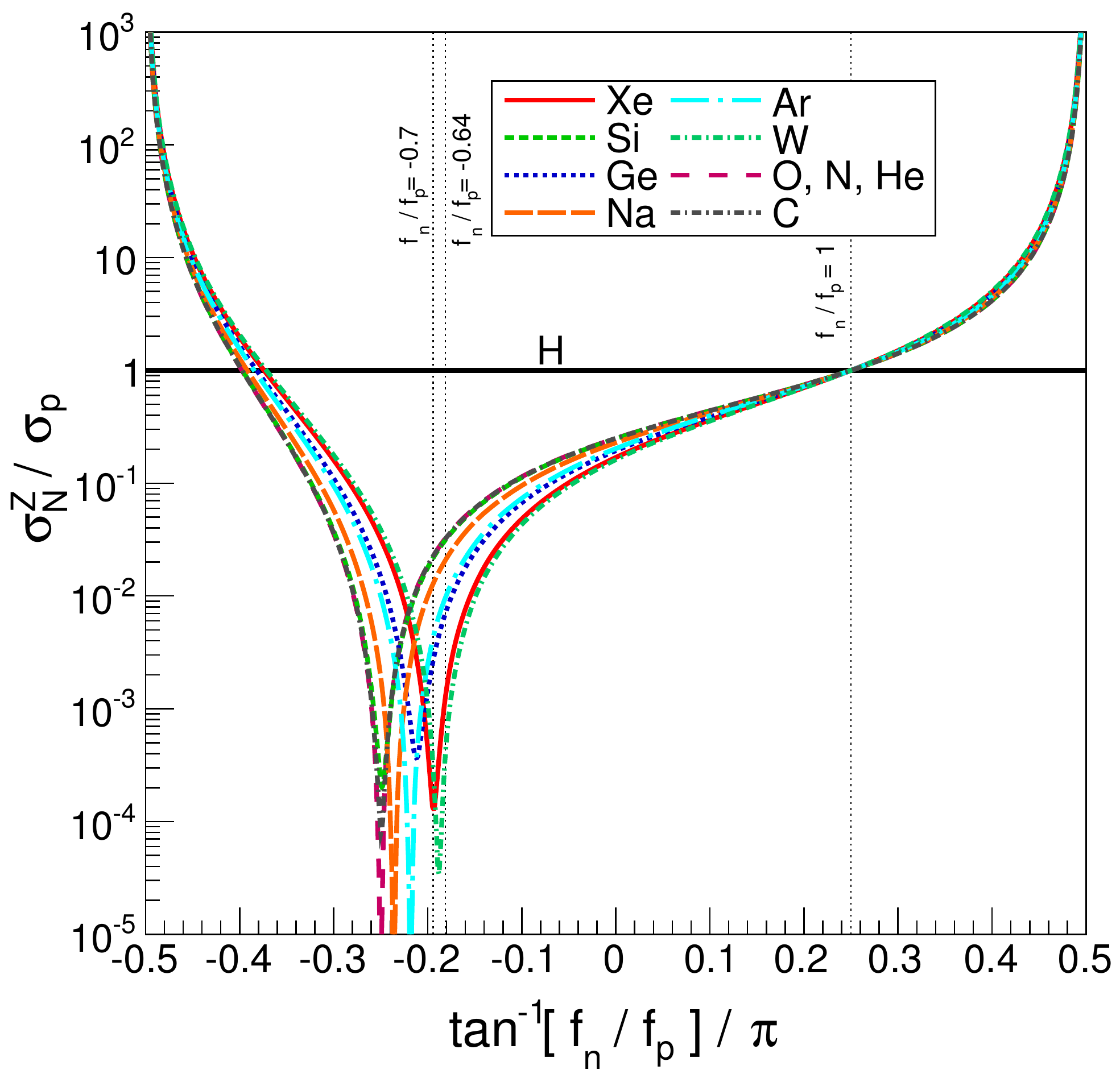}
\label{fig:FZp}}
\subfigure[\ $\sigma^N_Z / \sigma_n$, entire $f_n/f_p$ range]{
\includegraphics*[width=0.48\textwidth]{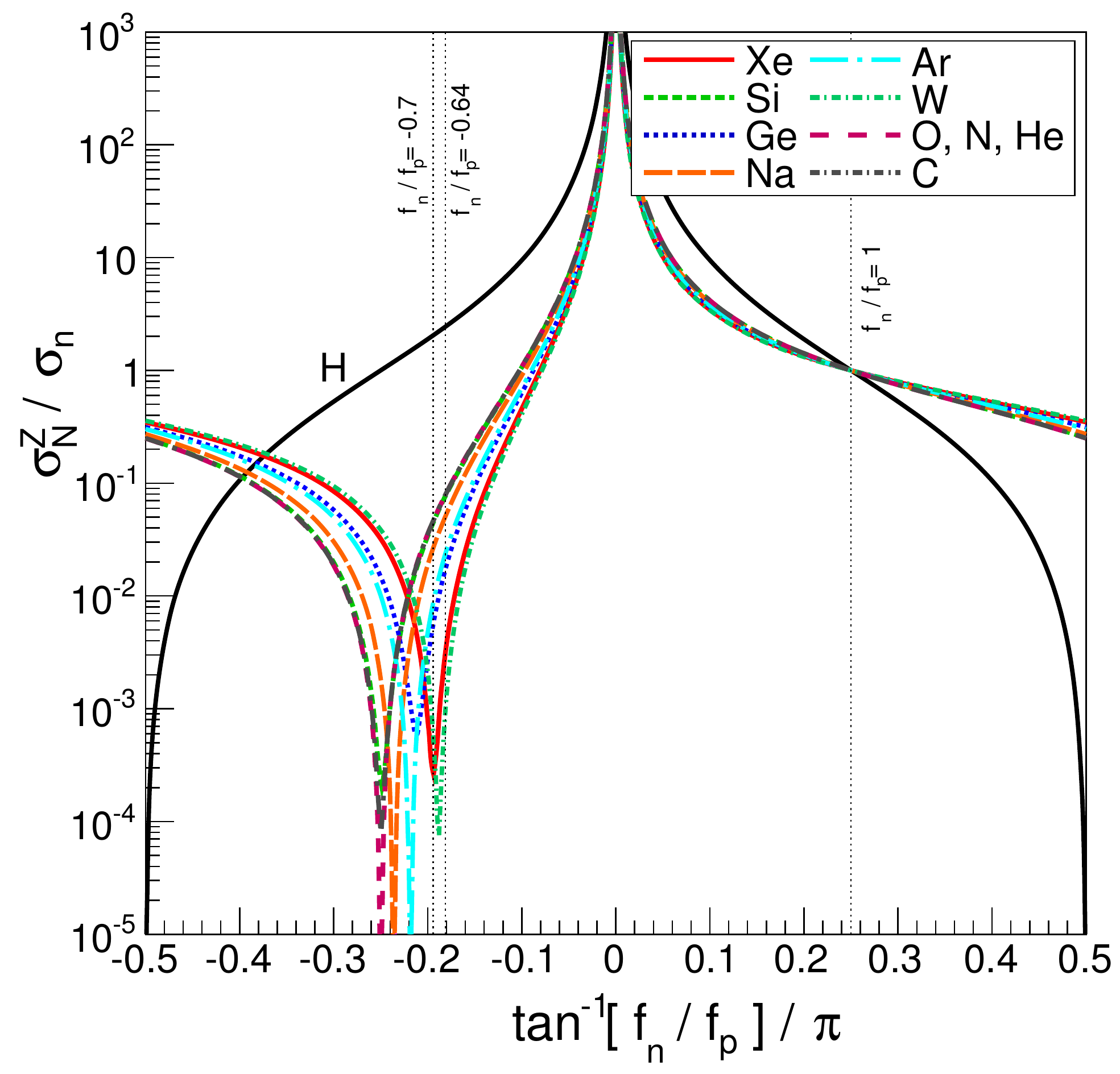}
\label{fig:FZn}}
\subfigure[\ $\sigma^N_Z / \sigma_p$, xenophobic region]{
\includegraphics*[width=0.48\textwidth]{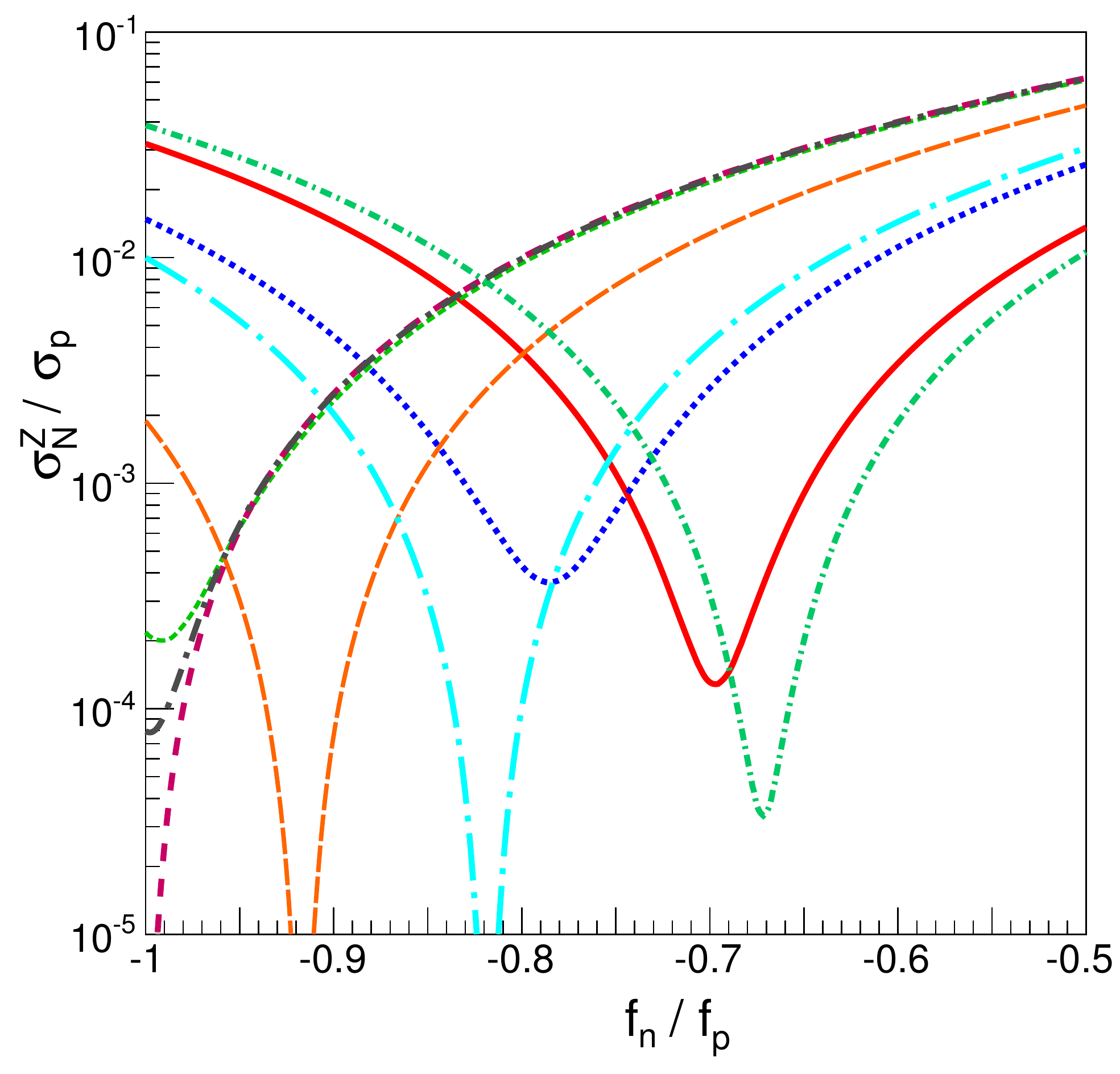}
\label{fig:FZp_small}}
\subfigure[\ $\sigma^N_Z / \sigma_n$, xenophobic region]{
\includegraphics*[width=0.48\textwidth]{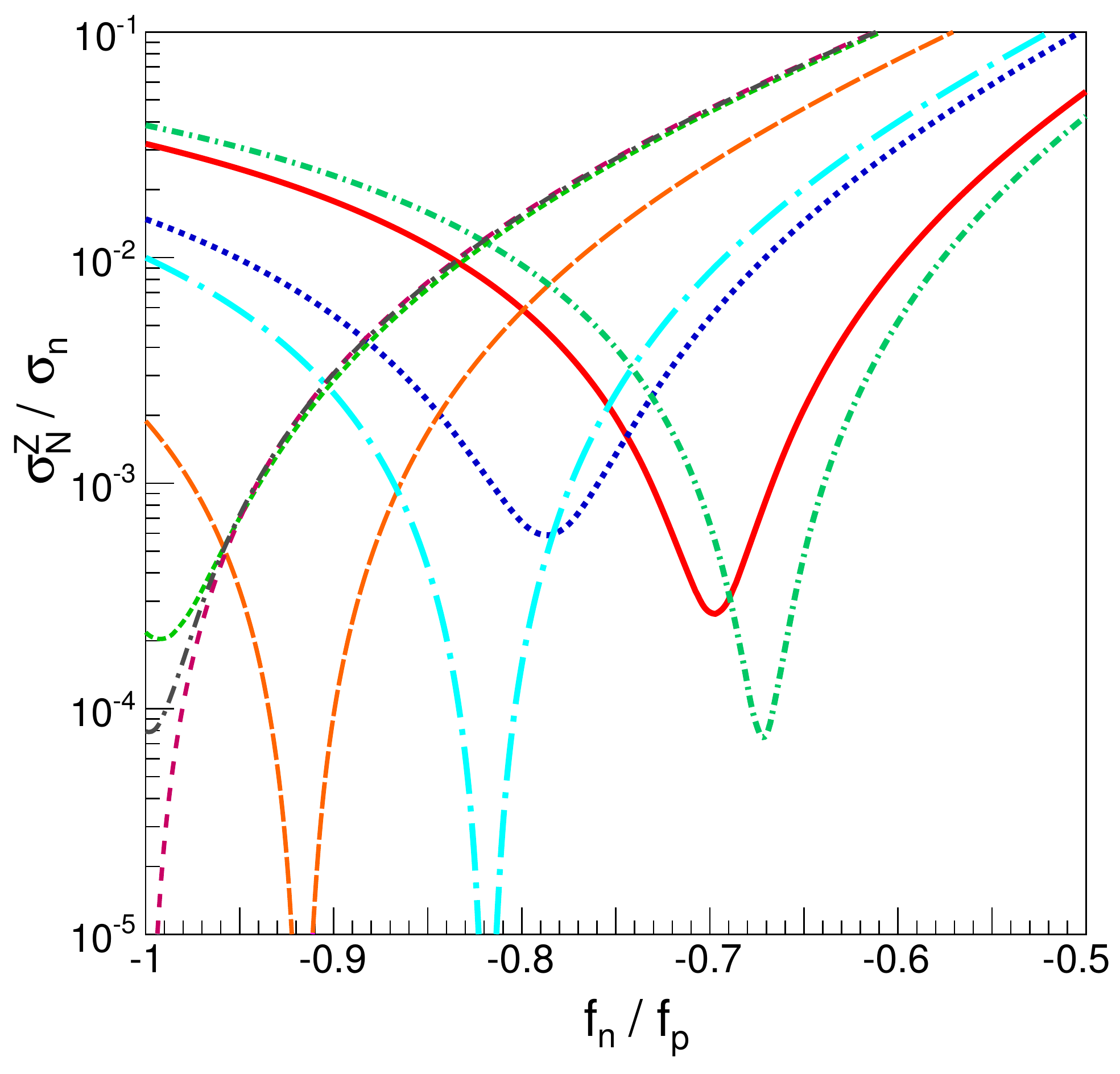}
\label{fig:FZn_small}}
\vspace*{-.1in}
\caption{\label{fig:FZ} \textit{Ratio of the normalized-to-nucleon
    cross section reported by direct detection experiments to the true
    nucleon cross section.}  Results are shown for $\sigma_N^Z /
  \sigma_p = D_p^Z$ (left) and $\sigma^N_Z / \sigma_n = D_n^Z$ (right)
  as a function of $f_n / f_p$ for various elements.  The entire range
  of $f_n / f_p$ is shown (top) as well as the xenophobic region
  (bottom).  All plots assume $m_X = 8~\gev$, but are highly
  insensitive to this choice.
}
\end{figure}

In \figref{FZ} we plot the degradation factors $D^Z_p$ and $D^Z_n$ as
a function of $f_n / f_p$ for various elements that are used as
targets for low-mass dark matter searches.  Generically the
sensitivity to $\sigma_p$ is reduced for $\left| f_n / f_p \right|
\rightarrow 0$ and enhanced for $\left| f_n / f_p \right| \rightarrow
\infty$, with the opposite behavior for sensitivity to $\sigma_n$.
However, in both cases, sensitivity is significantly reduced for $-1.5
\lesssim f_n / f_p \lesssim -0.5$ by destructive interference.  Nearly
complete destructive interference occurs for oxygen, nitrogen, helium,
sodium, and argon, each of which has only one isotope with significant
natural abundance; all other elements have a lower bound on the
reduction of sensitivity in the range of $3\times 10^{-5} - 5 \times
10^{-4}$.

\subsection{Astrophysical and Collider Probes}

If there is destructive interference between dark matter interactions
with protons and neutrons, then the dark matter-proton and dark
matter-neutron scattering cross sections, $\sigma_p$ and $\sigma_n$,
must both be larger than $\sigma_N^Z$ to keep fixed the cross section
for dark matter to scatter off the target atomic nucleus
(equivalently, $D_{p,n}^Z < 1$).  This implies an enhanced coupling to
up and down quarks, which in turn implies large potential signals from
dark matter annihilation to hadrons and from dark matter production in
conjunction with spectator jets or photons at colliders, such as the
LHC~\cite{Birkedal:2004xn,Feng:2005gj,Beltran:2010ww,Goodman:2010yf,%
  Fox:2011pm}.

To consider this possibility concretely, assumptions regarding the
type of interaction and couplings to each quark flavor are required.
Here we examine the case that dark matter interacts with standard
model quarks through a set of effective four-point contact operators.
We will consider the set of effective operators that contribute to a
spin-independent scattering matrix element (not suppressed by factors
of the relative velocity or momentum transfer) and to an $S$-wave
annihilation matrix element.  These operators are of interest because
they permit unsuppressed spin-independent scattering, which could
explain the low-mass direct detection data, and also permit $S$-wave
annihilation, which could provide signals at an indirect detection
experiment.

If dark matter is a spin-0 particle, then there is a unique such
contact operator of dimension 6 or less, ${\cal O}_S =(1/M_*) \phi^*
\phi \bar q q$.  If dark matter is a Dirac fermion, there is a
different such contact operator, ${\cal O}_D =(1/M_*^2) \bar X
\gamma^\mu X \bar q \gamma_\mu q$; there is no such operator if dark
matter is a Majorana fermion~\cite{Kumar:2011dr,Kumar:2013iva}.  For
either case, if we assume dark matter couples only to up and
down quarks, then a choice of $f_n / f_p$ uniquely fixes the relative
strength of the dark matter coupling to up and down quarks, and thus
uniquely fixes the contact operator up to an overall coefficient.
This is a conservative limit of the theory, as including non-zero
couplings to heavier quark flavors generically enhances both indirect
detection and collider signals, for a fixed direct detection signal.

For a fixed choice of interaction operator and $f_n / f_p$, one can
then translate bounds on the dark matter annihilation cross section
from an indirect detection experiment, or bounds on the $XX +
\text{jet}$ production rate at a collider, into bounds on the overall
coefficient of the effective contact operator.  This directly
corresponds to a bound on the spin-independent scattering cross
section.

Assuming dark matter annihilates only to up and down quarks, bounds on
the dark matter annihilation cross section~\cite{Kumar:2011dr} were
determined from stacked analyses of gamma-ray emissions from dwarf
spheroidal galaxies~\cite{Ackermann:2011wa,GeringerSameth:2011iw}.
Bounds on $\sigma_p$, as a function of $f_n / f_p$, were then
determined in Ref.~\cite{Kumar:2011dr}, and we will consider their
impact on xenophobic dark matter.  There exist systematic
uncertainties in the dark matter density profile of the dwarf
spheroidals that significantly impact these limits, possibly weakening
them by a factor of $\sim 2$ or strengthening them by a factor of
$\sim 10$.

Collider bounds were produced in Ref.~\cite{Kumar:2011dr} under the
same assumption that dark matter couples only to up and down quarks
through a contact operator that permits unsuppressed SI-scattering and
$S$-wave annihilation.  In that analysis, the number of $pp
\rightarrow XX+ \text{jet}$ events expected at the LHC was determined
in terms of the overall coefficient of the contact operator.  From a
comparison of the number of monojet events that passed the cuts to the
number expected from standard model background events, bounds on
$\sigma_p$ were determined as a function of $f_n / f_p$.

Here, we update this analysis using upper bounds on monojet events
from new physics at CMS with an integrated luminosity of 4.7
fb$^{-1}$~\cite{Chatrchyan:2012me}.  Signal events were generated
using MadGraph 5.1.5.9~\cite{Alwall:2011uj} with Pythia
6.4~\cite{Sjostrand:2006za} for showering and Delphes
2.0.5~\cite{Ovyn:2009tx} for detector emulation.  The analysis of
Ref.~\cite{Chatrchyan:2012me} provides bounds for monojet $p_T > 110$
GeV and four cuts on missing transverse energy at $\slashed{E}_T >
\{250, 300, 350, 400 \}$ GeV.  For the Dirac fermion case the
strongest limits on ${\cal O}_D$ come from the $\slashed{E}_T > 400$
GeV cut, while for the complex scalar case the strongest limits on
${\cal O}_S$ are produced by the $\slashed{E}_T > 350$ GeV cut.  We
will also consider the impact of these bounds on models of xenophobic
dark matter.

It is important to note, however, that these bounds arising from
indirect detection and monojet searches rely on two major assumptions:
it is assumed that dark matter interacts through a contact operator
even at the energy scales relevant for dark matter annihilation or
production, and that this operator permits $S$-wave annihilation.  If
dark matter interacts through a true contact operator, then the dark
matter scattering, annihilation and production matrix elements all
scale as $M_*^{-2}$, arising from the propagator of the exchanged
mediator in the limit where the energy scale of the process is much
smaller than the mediator mass.  However, if the energy $E$ or
momentum transfer scale $q$ of the process is larger than the mediator
mass, the matrix element will instead scale as $E^{-2}$ or $q^{-2}$.
This can suppress the dark matter annihilation matrix element ($E \sim
2m_X$) and production matrix element ($E \geq 2m_X$) relative to the
scattering matrix element ($E \ll q \sim 1-100~\mev$).  The
suppression can be substantial, around $(M_* / m_X)^4$ for mediator
masses lighter than the dark matter mass.

If a particular spin-independent direct detection signal is not
consistent with the Fermi bounds described here, the implication is
that the interaction cannot be mediated by a contact operator that
permits $S$-wave annihilation; it may instead be permitted by a
contact operator that permits $P$-wave annihilation, or the
interaction might not be realizable as a contact interaction at the
energy scales relevant for dark matter annihilation.  Similarly, if
collider monojet bounds are inconsistent with a particular direct
detection signal, the implication is that the interaction is mediated
by an interaction structure that is not a contact interaction at the
energy scales of the LHC.

\subsection{Multiple Experiments}

For a given experiment the physical quantity $\sigma_p$ is not truly
an observable quantity unless the experiment involves scattering of
dark matter off hydrogen -- it can only be inferred from $\sigma^Z_N$
using some assumption regarding the underlying theory.  However, one
may define the observable ratio
\begin{eqnarray}
R[Z_1, Z_2] \equiv {\sigma_N^{Z_1} \over \sigma_N^{Z_2}} = {D_p^{Z_1}
  \over D_p^{Z_2} }\ ,
\end{eqnarray}
which is the ratio of the normalized-to-nucleon scattering cross
sections that one would infer for the same dark matter candidate from
the data of detectors using two different target materials.  A
measured dark matter signal at two different experiments (using
targets with $Z_1$ and $Z_2$ protons, respectively) constitutes an
experimental measurement of $R[Z_1, Z_2]$.  But as we see from
\eqref{eq:FZ}, the equation $D_p^{Z_1} = R[Z_1, Z_2] \times D_p^{Z_2}$
is quadratic in $f_n / f_p$, with coefficients that are all determined
by atomic physics.  As a result, with a measurement of $R[Z_1, Z_2]$
from two different experiments with different targets, one can
determine $f_n / f_p$ up to a two-fold ambiguity.

\begin{figure}[tb]
\subfigure[\ Entire $f_n / f_p$ range]{
  \includegraphics*[width=0.48\textwidth]{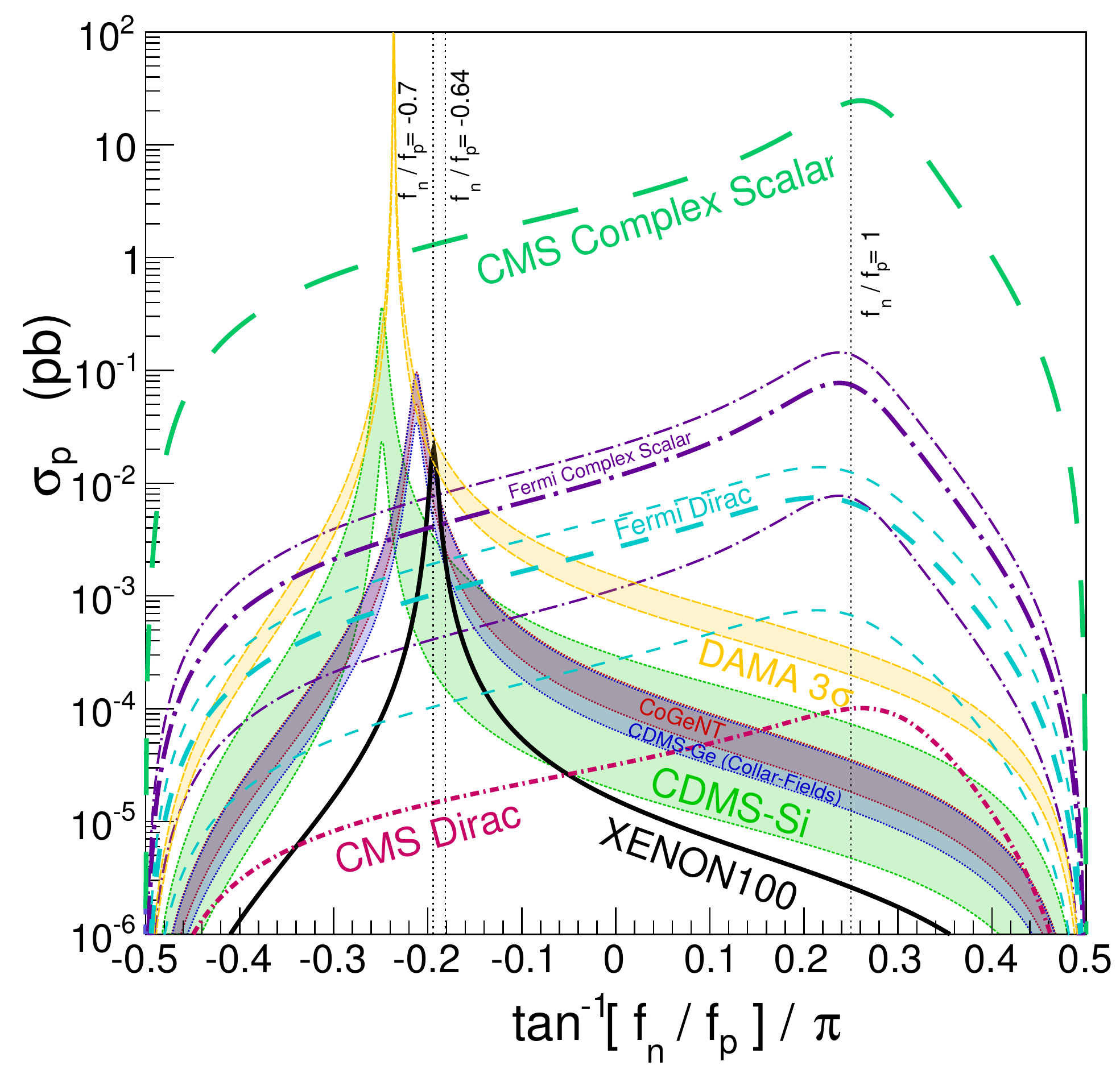}
\label{fig:sigmap_fnfp_angle}}
\subfigure[\ Xenophobic region]{
  \includegraphics*[width=0.48\textwidth]{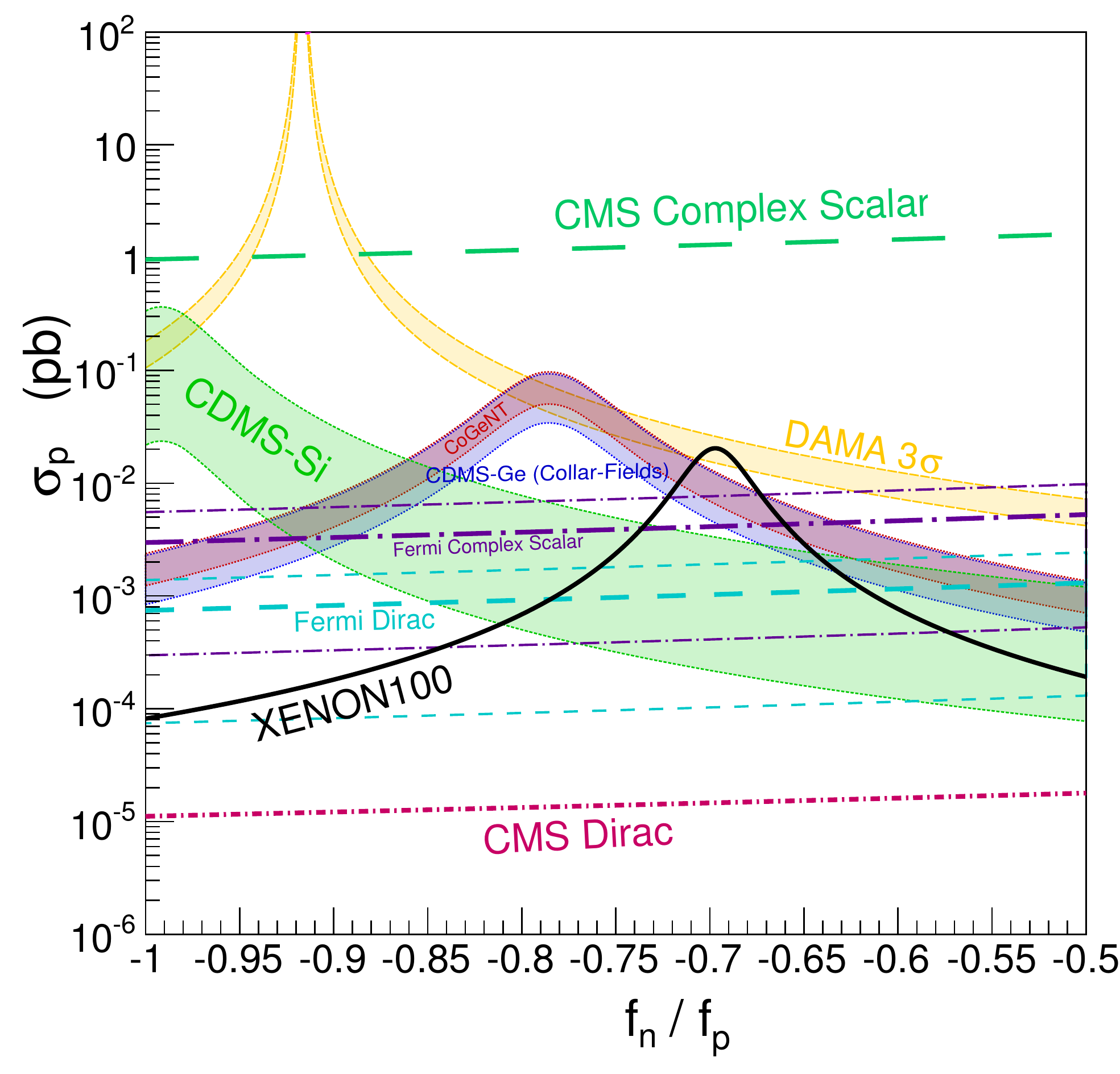}
\label{fig:sigmap_fnfp_linear}}
\vspace*{-.1in}
\caption{
\label{fig:sigmap_fnfp} \textit{Proton cross section for various
  experiments as a function of $f_n / f_p$ for $m_X = 8~\gev$.}
Plotted are slices of the 90\%~CL ROIs for
CDMS-Si~\cite{Agnese:2013rvf}, CoGeNT~\cite{Kelso:2011gd}, and CDMS-Ge
(Collar/Fields)~\cite{Collar:2012ed}, the $3\sigma$ ROI for
DAMA~\cite{Bernabei:2010mq}, and exclusion contours for
XENON100~\cite{Aprile:2012nq}.  Also plotted are 90\% CL exclusion
contours for CMS~\cite{Chatrchyan:2012me} and for the
Fermi-LAT~\cite{GeringerSameth:2011iw}, assuming dark matter is either
a complex scalar or Dirac fermion coupling only to first generation
quarks through an effective contact interaction permitting
unsuppressed spin-independent scattering and $S$-wave annihilation.
The thin dot-dashed violet and dashed teal lines correspond to the
systematic uncertainty in the Fermi-LAT bounds from astrophysical
uncertainties for complex scalar and Dirac fermion candidates,
respectively.}
\end{figure}

To illustrate this point, in \figref{sigmap_fnfp} we plot the range of
$\sigma_p$ that would be within the silicon- and germanium-based ROIs
at $m_X = 8~\gev$ as a function of $f_n / f_p$.  We also plot
exclusion contours from XENON100, from Fermi, and from CMS monojet
searches.  For the Fermi and CMS monojet search bounds, it is assumed
that dark matter is either a complex scalar or Dirac fermion that
interacts through a contact operator permitting $S$-wave annihilation
and spin-independent scattering with no momentum- or
velocity-suppression.  If dark matter is a real scalar, then the
Fermi-LAT bounds would be stronger than in the complex scalar case by
a factor of two, while the CMS monojet bounds would be weaker by a
factor of two~\cite{Kumar:2011dr}.

In particular, one sees that the $\sigma_p$ ROI corresponding to
CDMS-Si overlaps the germanium-based CoGeNT and CDMS ROIs for a wide
range of the parameter $f_n / f_p$, which includes the
isospin-invariant case $f_n / f_p =1$.  But there is another narrow
region, $f_n / f_p = -0.89 \pm 0.05$ for which the CDMS-Si and
germanium-based ROIs also overlap.  It is clear however that
XENON100's sensitivity relative to silicon or germanium based
experiments is enhanced in this second region, producing complete
exclusion.  Moreover, although Fermi would not probe models that could
match the silicon-based and germanium-based ROIs for $f_n / f_p =1$,
it rules out models that could match these regions for $f_n / f_p
\approx -0.89$, if dark matter interacts through a contact interaction
yielding $S$-wave annihilation.  Finally, we see from
\figref{sigmap_fnfp} that the silicon- and germanium-based ROIs
overlap yet again for $f_n / f_p \ll -1$.  The appearance of a third
overlap region may seem surprising, since $f_n / f_p$ is determined by
a quadratic equation.  But this result is readily understood from the
fact that the current regions of interest are of finite size.  Since
the silicon- and germanium-based ROIs are broad enough to be
consistent for $f_p \approx 0$, it is not surprising that the ROIs
overlap for both $f_n / f_p \gg 1$ and $f_n / f_p \ll -1$.  With
greater exposure of the detectors, the bands corresponding to these
ROIs should become thinner.  One can see from \figref{sigmap_fnfp}
that either of the overlap regions at $f_n / f_p \sim 1$ or $f_n / f_p
< -1$ could then disappear; indeed, one of these solutions would
necessarily go away.  However, the solution with $f_n / f_p \approx
-0.89$ is robust.

Although we have studied IVDM in the context of the particular details
of current low-mass data, the points we have made are quite generic.
In general, experimental signals of dark matter from two different
direct detection experiments can determine $f_n / f_p$ up to a twofold
ambiguity, which can be resolved by a detection or exclusion from a
third detector, and potentially by signals from indirect detection or
collider monojet searches.  The finite width of the ROIs supplements
need for at least three independent signals to determine $f_n / f_p$.

\section{Xenophobic Dark Matter}
\label{sec:xenophobic}

In the current generation of direct detection experiments the reported
sensitivity of XENON100~\cite{Aprile:2011hi,Aprile:2012nq} exceeds
that of all others by at least an order of magnitude, and the results
from the LUX experiment~\cite{Akerib:2012ak} are expected to exceed
that sensitivity significantly within the year.  From
\figref{sigmap_fnfp}, it is apparent that dark matter is maximally
xenophobic for $f_n / f_p \approx -0.70$ and the coupling
significantly suppressed for nearby values; however, for that value
the current silicon-based and germanium-based ROIs do not overlap.  On
the other hand, for slightly less xenophobic dark matter, $f_n / f_p
\approx -0.64$, the 90\% CL silicon- and germanium-based ROIs have a
region of overlap that is marginally consistent with exclusion
contours from XENON100.  For this choice of $f_n / f_p$, we plot in
\figref{sigmap_-64} the silicon- and germanium-based ROIs, and
XENON100, Fermi and CMS monojet exclusion contours as a function of
$m_X$, along with projected limits from LUX.  We thus see that, though
this region of parameter space can potentially reconcile the current
germanium-, silicon-, and xenon-based detector data, it can be
decisively probed if data from LUX significantly improves upon
XENON100's current sensitivity.  The current projected sensitivity at
LUX does not conclusively probe the disputed region, and indeed the
LUX experiment claims no sensitivity to dark matter with $m_X \lesssim
7~\gev$; however, the LUX collaboration uses very conservative
estimates for their light collection efficiency, and a dedicated
ionization-only analysis could still produce sensitivity to the
low-mass region\cite{Hooper:2013cwa}.

It is interesting to note that this model is in tension with both
collider and Fermi bounds if dark matter is a Dirac fermion
interacting through a contact operator that permits $S$-wave
annihilation.  However, if dark matter is a complex scalar that
couples through an effective contact operator permitting $S$-wave
annihilation, then this model is consistent with collider bounds, but
only marginally consistent with bounds from Fermi searches of dwarf
spheroidal galaxies.  In particular, systematic uncertainties in the
dark matter density profile of the dwarf spheroidals can have a large
impact on the consistency of the Fermi data with this model.  This
suggests that any future results from Fermi, positive or negative,
could have an interesting impact on this scenario.  If indeed the data
is explained by a model in which dark matter is a xenophobic complex
scalar interacting through an effective contact operator, then one
should expect that Fermi will soon see an excess of gamma-rays from
dwarf spheroidal galaxies.  If Fermi does not see such an excess, it
suggests that a model of this type can only be consistent with the
data if the interaction is not a contact interaction.  If dark matter
interacts through a contact operator that only permits $P$-wave
annihilation, then although the Fermi bounds would be satisfied, the
bounds from collider searches would become problematic.

\begin{figure}[tb]
\subfigure[\ Entire $f_n / f_p$ rangle]{
\includegraphics*[width=0.48\textwidth]{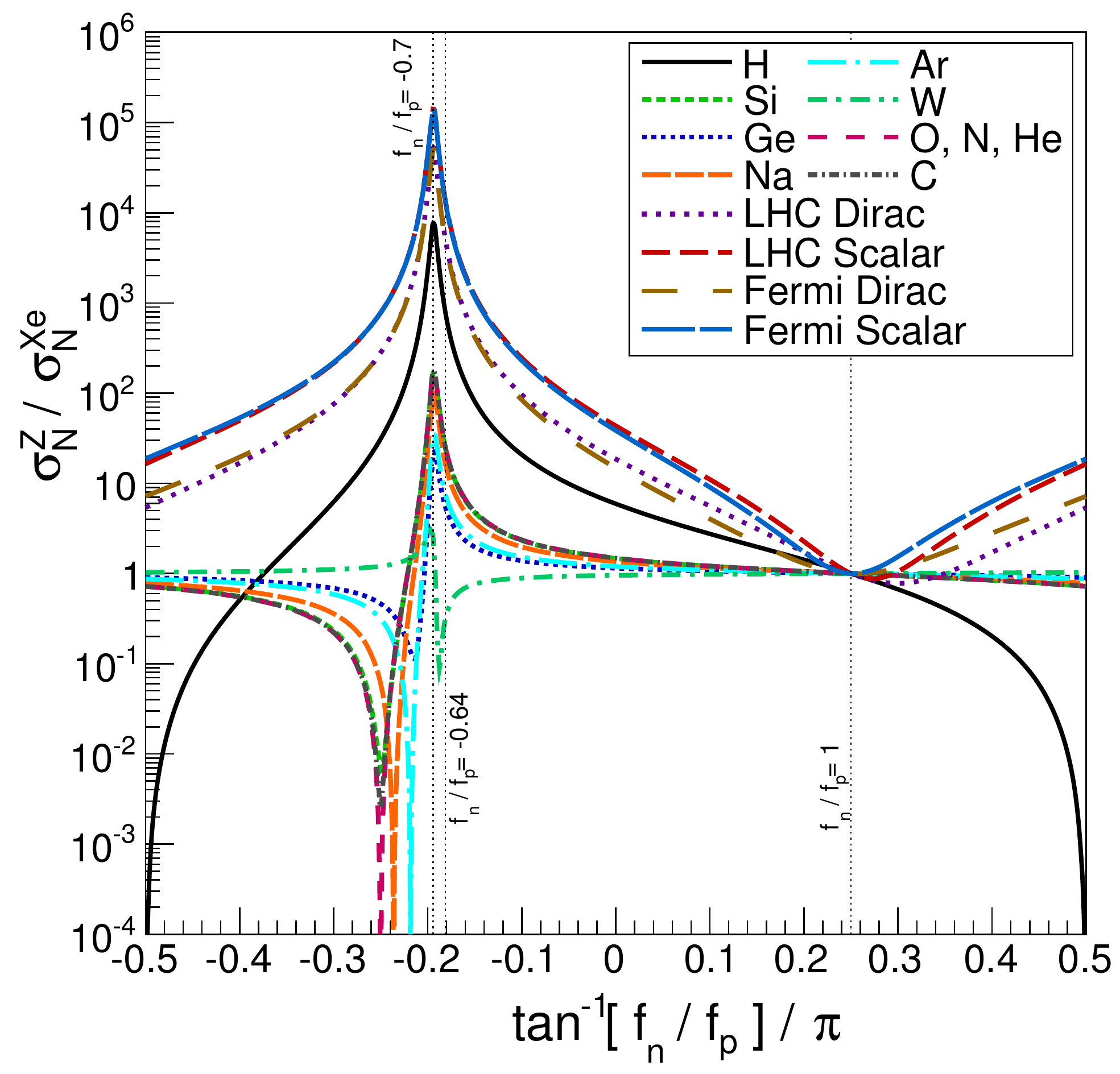}
\label{fig:xenon_ratio_angle}}
\subfigure[\ Xenophobic region]{
  \includegraphics*[width=0.48\textwidth]{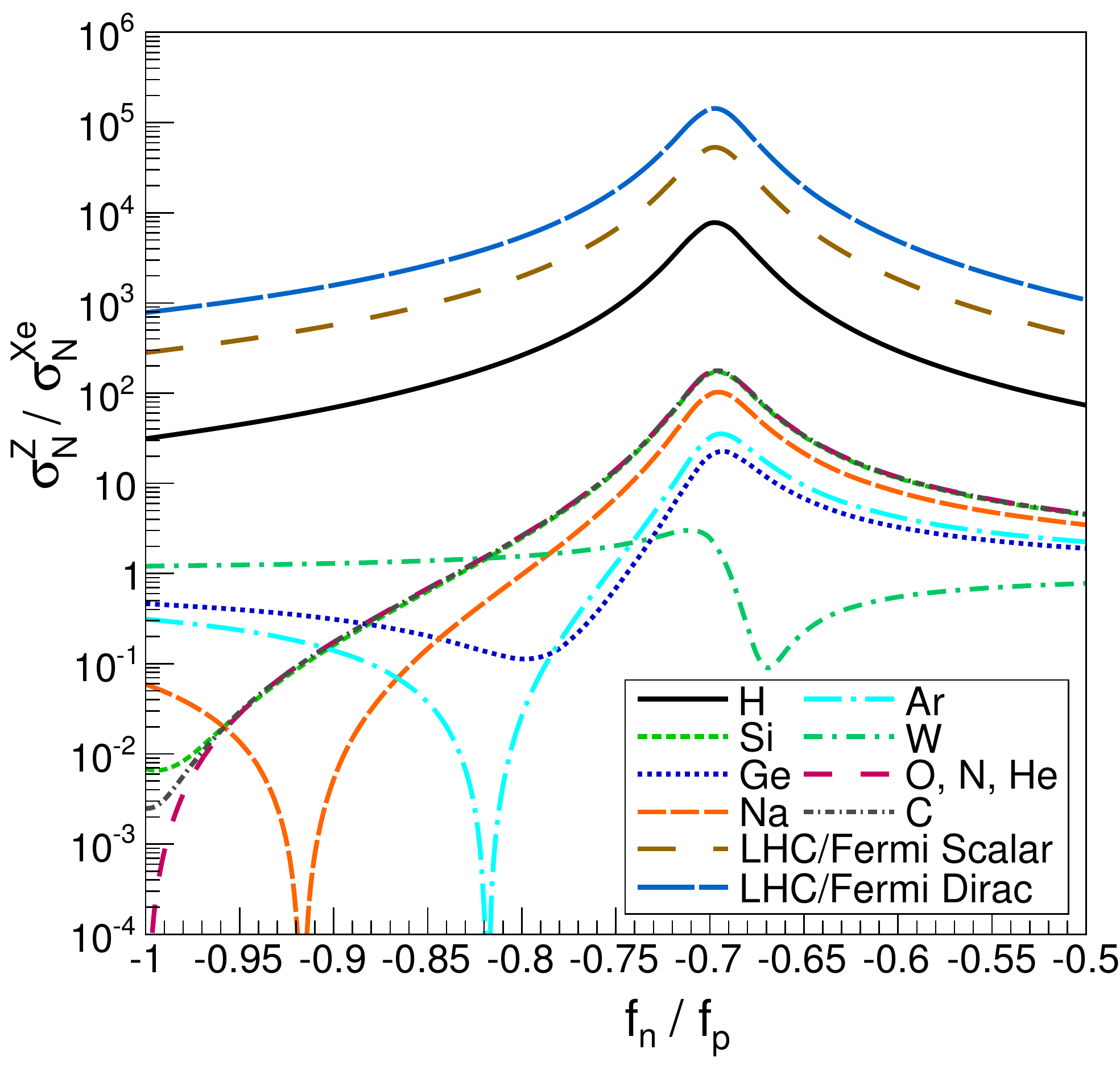}
\label{fig:xenon_ratio_linear}}
\vspace*{-.1in}
\caption{\label{fig:xenon_ratio} \textit{Ratio of $\sigma^Z_N$ in
    various experiments to $\sigma^{Xe}_N$.}  Results are shown as a
  function of $f_n / f_p$ for scattering off various elements, as well
  as for LHC and Fermi determinations.  In the xenophobic region the
  behavior of LHC and Fermi bounds for a given operator are visually
  identical.}
\end{figure}

The overall enhancement of various experimental signals relative to
xenon-based detectors is shown in \figref{xenon_ratio}.  Although
$D_p^{\text{Xe}} \sim 10^{-4}$ at its minimum, as shown in
\figref{FZ}, $D_p^Z$ is also suppressed for all elements except
hydrogen in the range $-1.5 \lesssim f_n/f_p \lesssim -0.5$.  As a
result, the maximal value of $R[Z, \text{Xe}]$ ranges from $\sim 20$
to $\sim 200$ for various lighter elements relevant for direct
detection.  In contrast, collider and annihilation signals suffer no
suppression in this region, and are even enhanced relative to
scattering off protons, resulting in a maximal $R[\{\mathrm{LHC,
    annihilation}\}, \text{Xe}]$ of $\sim
10^5$.~\footnote{$R[\{\mathrm{LHC, annihilation}\}, \text{Xe}] \equiv
  \sigma_N^{\{\mathrm{LHC, annihilation}\}} / \sigma_N^{Xe}$, where
  $\sigma_N^{\{\mathrm{LHC, annihilation}\}}$ is the dark
  matter-nucleon scattering cross-section that would be inferred from
  LHC/indirect detection data if one assumed $f_n / f_p =1$.}  It is
also worth noting that, as one moves away from the maximally
xenophobic limit, one would expect NLO corrections to the scattering
cross section to be less important.

\section{Conclusions}
\label{sec:conclusions}

We have revisited the discussion of isospin-violating dark matter as a
way of potentially reconciling several recent positive signals at low
mass from direct detection experiments with very tight exclusion
contours from xenon-based detectors.  Our focus has been on xenophobic
dark matter: dark matter in which destructive interference between
coupling to protons and neutrons drastically reduces the sensitivity
of xenon-based detectors.  We note, importantly, that the large
natural abundance of several xenon isotopes implies that even
xenophobia has its limits~\cite{Feng:2011vu}; there is no choice of
parameters that can completely eliminate the response of all xenon
isotopes.

Focusing on recent positive signals from CDMS-Si detectors and the
CoGeNT experiment, and on a ROI identified by an analysis of CDMS-Ge
detectors from Collar and Fields, we have found that these ROIs can
potentially be made to overlap in a regions marginally consistent with
bounds from XENON100 for dark matter that is near maximally
xenophobic, with $f_n / f_p \approx -0.64$.  While a true global
likelihood analysis to determine if this region is a good fit to the
combined data is beyond the scope of this work, even in this
prescription the improvement in consistency is qualitatively clear.

Moreover, we have only focused on the effect of isospin-violation;
changes to astrophysics assumptions can alter this picture, possibly
producing more alignment of current results.  New results will also
alter the picture, in particular new data from CoGeNT that may result
in refining their ROI.  More generally, we have shown that the results
from multiple detectors and from independent detection strategies,
such as indirect or collider searches, provide important complementary
data, which are necessary for clarifying the consistency of the
low-mass data.  In particular, new results from LUX and from Fermi
dwarf spheroidal searches should provide important tools for testing
models of xenophobic dark matter.

This analysis highlights the importance of improvements in direct
detection experiments for clarifying the viability of models of
low-mass dark matter.  In particular, even though it is a xenon
detector, LUX may have much to say about the xenophobic models
discussed here.  This hinges critically on LUX's sensitivity to $\sim
8~\gev$ dark matter, which will depend in detail on the charge and
light yields of liquid xenon (as well as the backgrounds) at low
recoil energies.  LUX may be capable of achieving a low-mass
sensitivity significantly greater than the estimates used here, but
such an assessment must likely await a full analysis of the data.

\section*{Acknowledgments}

We are grateful to E.~Figueroa and D.~McKinsey for useful discussions.
We are grateful to the organizers of the APS April Meeting, to the
organizers of the Light Dark Matter Workshop at MCTP, and to the KITP,
where part of this work was conducted.  The work of JLF is supported
in part by NSF Grant No.~PHY--0970173 and a Simons Foundation
Fellowship.  The work of JK prior to June 1, 2013 was supported in part by
U.~S.~Department of Energy grant DE--FG02--04ER41291.  DS is supported
in part by U.S.~Department of Energy grant DE--FG02--92ER40701 and by
the Gordon and Betty Moore Foundation through Grant No.~776 to the
Caltech Moore Center for Theoretical Cosmology and Physics.

\bibliography{bibxenophobic1e}{}

\end{document}